\documentclass[prl,aps,ams,twocolumn,epsfig,superscriptaddress]{revtex4-1}
\usepackage{epsfig}
\usepackage{amsmath}
\usepackage{amssymb}
\usepackage{bm}
\usepackage{dcolumn}% Align table columns on decimal point

\begin{document}

\title{Trial wavefunctions for the Goldstone mode in $\nu=1/2+1/2$ quantum Hall bilayers}

\author{Gunnar M\"oller}
\affiliation{TCM Group, Cavendish Laboratory, JJ Thomson Avenue, Cambridge CB3 0HE, UK}

\author{Steven H. Simon}
\affiliation{Rudolf Peierls Centre for Theoretical Physics, Oxford, OX1 3NP, UK}

\begin{abstract}
Based on the known physics of the excitonic superfluid or 111 state of the quantum Hall $\nu=1/2+1/2$ bilayer, we create a simple trial wavefunction ansatz for 
constructing a low energy branch of (Goldstone) excitations by taking the overall ground state and boosting one layer with respect to the other.  
This ansatz works extremely well for any interlayer spacing.  For small $d$ this is simply the physics of the Goldstone mode, whereas for large $d$ this is a 
reflection of composite fermion physics.  We find hints that certain aspects of composite fermion physics persist to low $d$ whereas certain aspects of 
Goldstone mode physics persist to high $d$.   Using these results we show nonmonotonic behavior of the Goldstone mode velocity as a function of $d$. 
\end{abstract}
\date{August 1, 2010}
%\pacs{}
\maketitle

The $\nu=1/2+1/2$ quantum Hall bilayer is a remarkably rich system \cite{DasSarmaPerspectives,DLReviewNature}.
At small enough spacing between the layers, $d$, the system is
known to be an excitonic superfluid \cite{MoonetalPRB95} known sometimes as the 111 phase \cite{Halperin111}.
At larger layer spacing, a phase transition or crossover is observed experimentally \cite{MurphyPRL94,PRL84Spielmanetal00,Tutucetal04,
Wiersmaetal04,Kellogg04,Kumadaetal05,Champage08} leading
to a compressible phase which is well described by two weakly coupled composite fermion Fermi liquids.   The nature of this crossover, and whether there are
intervening phases between small and large $d$, has been a matter of some debate in the community \cite{MoonetalPRB96, JoglekarPRB01,KimNayak01,
SchliemannPRL01, PRLSimon03, Yoshioka06, MollerSimonRezayiPRL, MollerSimonRezayiPRB}. 
%The understanding of this system is further complicated by the role of disorder for transport .

There are some results, however, that are extremely well established theoretically.  In the limit where $d$ becomes very small, it is known that the Halperin 111
trial wavefunction becomes exact \cite{Halperin111}. In a more BCS-like language this wavefunction can be expressed as \cite{MoonetalPRB95}
\begin{equation}
\label{eq:111}
  |111\rangle = \prod_X (c^\dagger_{X,\uparrow} + c^\dagger_{X,\downarrow}) | 0 \rangle
\end{equation}
where $\uparrow$ and $\downarrow$ indicate the layer index (we assume the real spin is frozen throughout this paper) and  $X$ constitutes the orbital index within
the Lowest Landau level (chosen to be the $x$-directed momentum in Landau gauge for example)\footnote{Strictly speaking this second quantized form of the
wavefunction must be projected to fixed number of particles within each layer to generate a Halperin 111 wavefunction.  However, in the thermodynamic limit these
two descriptions are essentially equivalent}.

The BCS-like form of Eq. \ref{eq:111} allows one to consider long wavelength Goldstone excitations of the form \cite{Rasolt85,MoonetalPRB95}
\begin{equation}
\label{eq:111_excited}
  |111-\text{excitation},\, k \rangle = \prod_X (c^\dagger_{X,\uparrow} + e^{i k X} c^\dagger_{X,\downarrow}) | 0 \rangle
\end{equation}
These modes are expected to form a linearly dispersing low energy branch with energy proportional to $k$  for small $k$. Physically, this Goldstone mode
corresponds to superflow --- one layer being boosted with respect to the other. Both a linearly dispersing mode \cite{SpielmanetalLinear} and excitonic superflow
\cite{PRL88Kelloggetal02,PRL90Kelloggetal03,Kellogg04,Tutucetal04,TutucShayegan05} were observed experimentally in this system.

Away from the $d \rightarrow 0$ limit the form of the bilayer ground state is not known exactly.   However, so long as we remain in the same phase of matter, there
will continue to be a linearly dispersing Goldstone mode in the long wavelength limit. An approximate expression for this Goldstone mode can be obtained from the
ground state wavefunction at any $d$ simply by boosting one layer with respect to the other.   One purpose of the current paper is to test this technique of
generating trial wavefunctions for the long wavelength Goldstone modes.

We note that this technique is not expected to be exact away from $d=0$, but for small $d$ is expected to be quite accurate.   In a conventional picture of
superfluidity, one might imagine that it would be better to find a way to boost the superfluid fraction while leaving the ``normal" fraction unboosted  (only at $d=0$ is
the system entirely super in some sense \cite{JoglekarPRB01}).   Nonetheless, our technique appears to work quite well even
away from $d=0$.

One might expect that once the system is no longer in the 111 phase of matter (roughly $d > 1.5$ magnetic lengths), our technique for generating excited states
would fail.  However, this turns out not to be the case.   First of all, at intermediate $d$ there may exist an interlayer paired state as discussed in
Ref.~\onlinecite{MollerSimonRezayiPRL}.  Such a paired state would also have a Goldstone mode that could be generated from the ground state by boosting one
layer in exactly the same way.

However, even at very large $d$ when such a pairing phase is either absent or pairing is extremely weak, our scheme for generating excited states still works
surprisingly well.  To understand why this is so, we realize that at large enough $d$ each layer is essentially independent.    To first approximation, each layer
forms a composite fermion Fermi liquid, which for finite size system has finite momentum (except when the number of electrons exactly fills a shell).  The two Fermi
liquids are weakly coupled and can combine their momenta to form an overall zero momentum ground state.  But since the coupling between the two layers is
weak, it costs very little energy to instead form a state of overall finite momentum --- which can be interpreted as boosting one layer with respect to the other in
comparison to the ground state.

\begin{figure*}[ttbp]
  \begin{center}
        \includegraphics[width=0.78\textwidth]{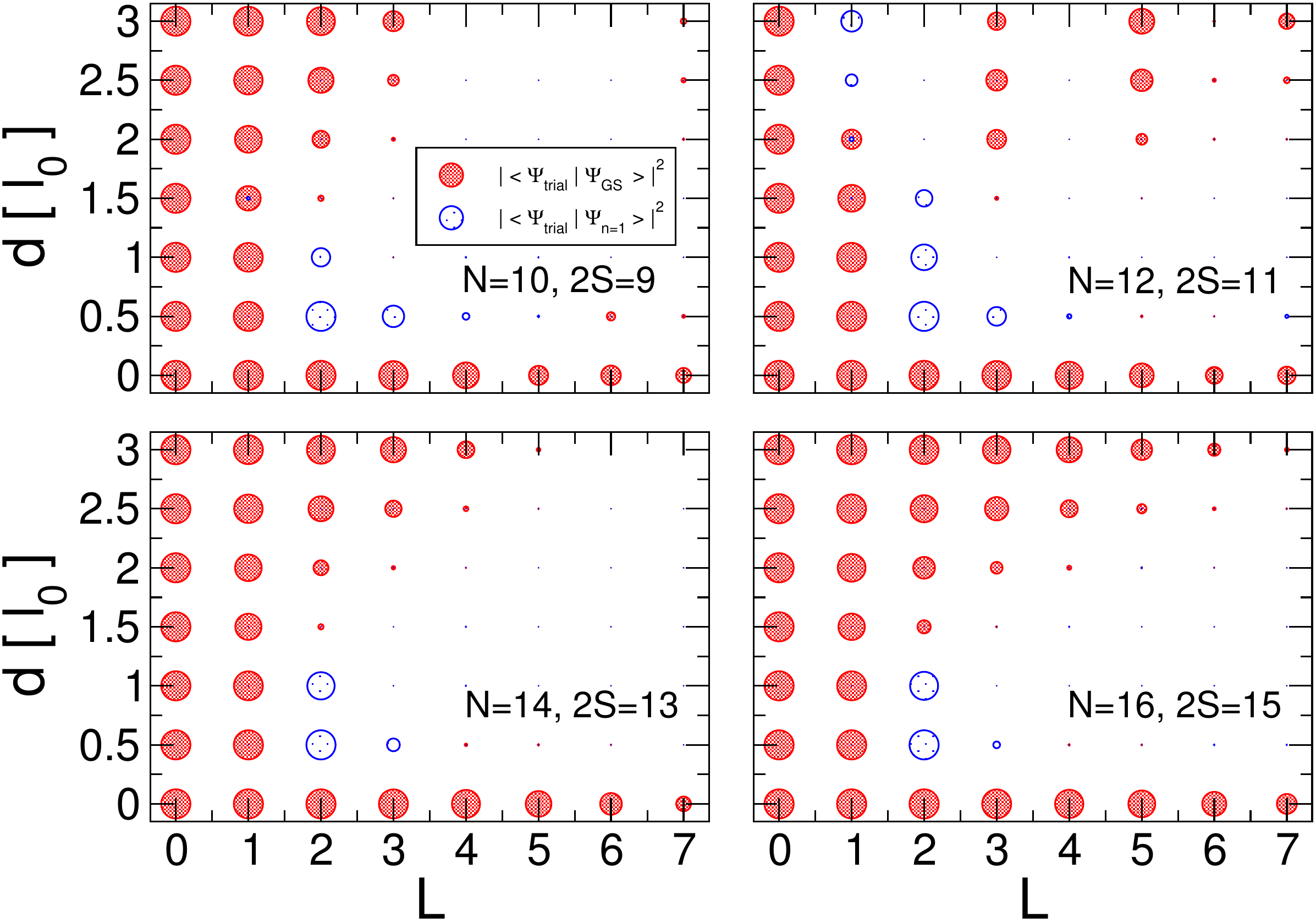}
  \end{center}
\caption{  \label{fig:overlaps} Overlaps of the trial states (\ref{eq:TrialState}) with the exact eigenstates of the Coulomb Hamiltonian for a selection of
system sizes $N=10$, $12$, $14$ and $16$ on the sphere. The magnitude of the overlap of the trial state with the exact groundstate is indicated by the size of red
dark shaded circles. Blue lightly shaded circles additionally indicate the overlap of the same trial states with the first excited state of the exact spectrum. Overlaps
at $L=0$ are equal to one by definition and give the overall scale. The trial states are very accurate at $d=0$, giving a good description of the lowest energy mode in each
sector of angular momentum (the Goldstone mode) up to large $L$. At finite layer separation $d$, the Goldstone mode is always present at small $L$, but does not reach to similarly
high values of angular momentum. The description is again more successful at very large $d$.
}
\end{figure*}

In the absence of any superfluid order parameter (at large $d$) it is probably not strictly appropriate to refer to this low energy mode as a Goldstone mode.  However, since this mode may evolve continuously into the Goldstone mode at smaller $d$ we will abuse nomenclature and continue to call it a Goldstone mode.  (If, as conjectured in Ref.~\onlinecite{MollerSimonRezayiPRL}, the bilayer is actually paired out to large $d$, then the usage remains correct).

Throughout this paper we will work with a spherical geometry.  In this case, boosting one layer with respect to the other corresponds to applying the angular
momentum raising operator $L_+$ to one layer but not the other  (call this operator $L_{+, \uparrow}$ meaning that it is applied to the $\uparrow$ layer only).
In the appendix we show that if we start with any $L=0$ state of the entire system, applying $(L_{+,\uparrow})^J$ generates a bilayer state with overall angular
momentum $L=L_z=J$.   Our technique is then to use exact diagonalization to generate the $L=0$ ground state of the bilayer system, which is used to obtain
the trial wavefunction for the excited state
\begin{equation}
\label{eq:TrialState}
  |\,\text{Trial}(d): L=J\rangle =  (L_{+,\uparrow})^J  \;|\,\text{Ground State}(d) : L=0 \rangle.
\end{equation}
In turn, we compare this trial state to the exact excited states with angular momentum $L=J$.

Our numerical work is based on exact diagonalization of the Coulomb Hamiltonian for a bilayer system on the sphere \cite{PRLHaldane83}. We simplify the problem
to exclude issues related to a finite tunneling amplitude between the layers, Landau-level mixing, or spin (which we assume is polarized), and model each layer as an ideal 2D plane without
considering its width into the third dimension. At fillings smaller than one per layer, the Hamiltonian is thus given by the projection of the Coulomb interaction
into the lowest Landau-level
\begin{equation}
\mathcal{H}[d] = \!\sum_{ \sigma = \uparrow,\downarrow \atop i<j }
 \frac{e^2}{\epsilon | r_{\sigma,i} - r_{\sigma,j}|}
 + \sum_{i,j} \frac{e^2}{\epsilon \sqrt{ |r_{\uparrow,i} - r_{\downarrow,j}|^2 + d^2}},\!
\end{equation}
where sums run over all particles with the given pseudo-spin. The interactions are parametrized by the layer separation $d$ that is measured in units of the
magnetic length $\ell_0=\sqrt{\hbar c/eB}$. All lengths given in this paper should be understood to be measured in units of $\ell_0$, where this is not explicitly
indicated.

\begin{figure}[tttt]
  \begin{center}
    \includegraphics[width=0.85\columnwidth]{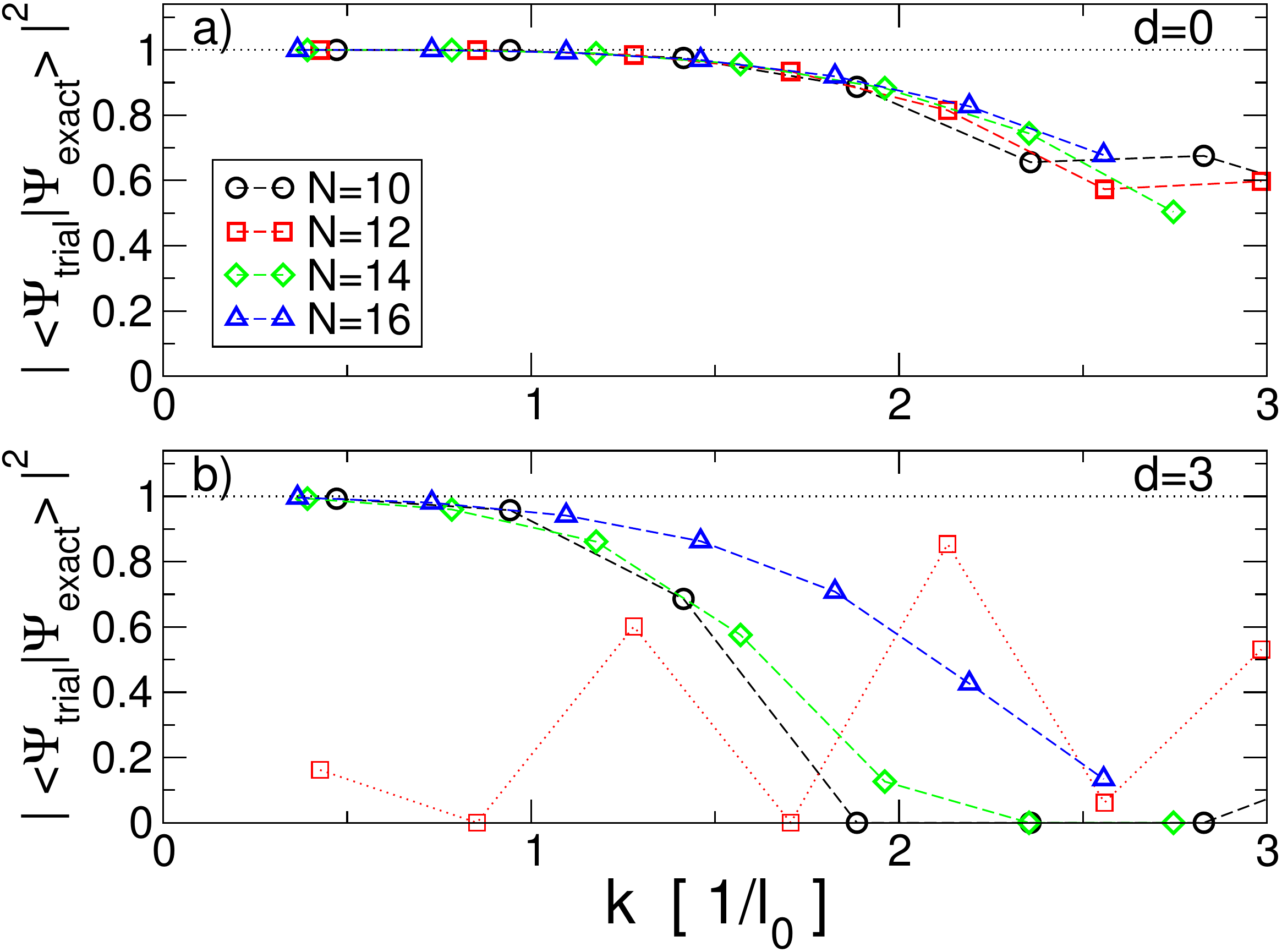}
  \end{center}
\caption{  \label{fig:overlaps_d_0} Overlaps of the trial states for the Goldstone mode at $d=0$ (top) and $d=3$ (bottom) with the exact state,
as in Fig.~\ref{fig:overlaps}, plotted here as a function of
wavevector $k \sim L/\sqrt{N_\phi/2}$. Despite the size of the Hilbert spaces increasing significantly between the smallest and largest system shown, the
overlap remains roughly constant or maybe slightly increases with $N$. The region of high overlap extends roughly up to $k\approx 2 \ell_0^{-1}$.  The failure of this approach for $d=3$ and $N=12$ is discussed in detail in the text.
}
\end{figure}

In our exact diagonalization calculations, we focus specifically on the density-balanced bilayer system with $N_\uparrow=N_\downarrow=N/2$ and consider the
state at the shift of the $111$ state, namely $N_\phi=N-1$. We obtain the two lowest-lying eigenvalues and eigenvectors in each sector of angular momentum.
This is most easily achieved using a projected Lanczos algorithm \cite{WojsPRL} which uses an additional projection to the
lowest energy subspace of minimal angular momentum after each multiplication with the Hamiltonian.
In a given sector with fixed $L_z$, this procedure therefore directly yields eigenstates of $L^2$ with the eigenvalue $L=L_z$.

To assess the accuracy of the trial states (\ref{eq:TrialState}) for the Goldstone mode, we consider their overlap with the lowest energy state in each sector of
angular momentum $L^2$ and for layer separations $d=0\ldots 3\ell_0$, in steps of $\frac{1}{2}\ell_0$ 
\footnote{We will attempt keep to the convention of using the word ``groundstate'' to indicate the absolute lowest energy state of the system, whereas the lowest energy state of an angular momentum sector will be referred to as such.}.
We note again that the the trial states are generated by applying the operator $(L_{+,\uparrow})^J$ to the exact groundstate at $L=0$. Only at $d=0$ is
an exact analytical expression of the groundstate known -- the $111$-state.  At other values of $d$, the numerical groundstates from exact diagonalization
are used, although very accurate trial wavefunctions are also known \cite{MollerSimonRezayiPRB}.
The results are summarized in Fig.~\ref{fig:overlaps}, which also indicates overlaps with the first excited state in addition to the overlaps with lowest energy state
in each sector. At $d=0$, the ansatz (\ref{eq:TrialState}) is very successful, describing excited states up to high angular momentum accurately.
This is shown in more detail in Fig.~\ref{fig:overlaps_d_0} a), which displays the magnitude of the overlap as a function of wavevector $k$. Surprisingly, for $d=0$
the overlap is very consistent with system size at given $k$, even though the Hilbert space dimension increases strongly with $N$. A very good 
description with overlaps
above $0.8$ is given up to a wavevector of $k\sim 2\ell_0^{-1}$. Turning back to Fig.~\ref{fig:overlaps}, we now focus on the overlaps at finite values of
layer separation. At $d=0.5\ell_0$, our trial states obtain significant overlaps only with the first excited state at $L=1$. This is not due to a disappearance of the
linearly dispersing mode however. Rather, a level crossing appears with distinct excitations occurring at energies less than that of the Goldstone mode,
as can be seen from the significant overlaps with the first excited excited state in the sector of $L=2$. At $d>1.5\ell_0$ this overlap with the first excited state
again disappears, signalling the presence of additional low-lying excitations of a nature different from the linearly dispersing Goldstone mode.

Finally at the largest value of $d=3\ell_0$, the ansatz for the boosted trial wavefunctions becomes more accurate than at intermediate $d$, signalling the
possible emergence of a distinct mode of low-lying excitations. Based on the overlap with the trial states, we can point out that there are strong
finite size effects in the physics at large $d$. As with $d=0$, Fig.~\ref{fig:overlaps_d_0} b) displays the numerical values of the overlaps at $d=3$. These
data single out the system with $N=12$ particles as particularly poorly described by these trial states. Here, the physics at large $d$ is clearly dominated by the shell filling effects of composite fermions.  As $N_\uparrow=N_\downarrow=6$ electrons per layer precisely fill the lowest two shells of composite fermion orbitals in one quantum of effective flux, this system size is
aliased with the situation where each layer forms its own incompressible $\nu=2/5$ state with angular momentum zero in each layer ---  making it impossible to form higher angular momenta states by boosting one layer with respect to each other.

Before proceeding further, note the rather unusual feature that, excepting $N=12$, the trial states give higher overlaps for larger systems at $d=3\ell_0$.
This unusual behaviour is related to a different manifestation of the composite fermion 
shell filling effect. As we have shown in Ref.~\onlinecite{MollerSimonRezayiPRL}, the
\emph{groundstate} at large layer separation is a state in which each layer individually obeys Hund's rule and maximizes the angular momentum per layer 
$L_\sigma=L_\uparrow=L_\downarrow$, while both layers are combined into a total $L=0$ state. Without modifying the correlations inside each layer, the same states 
with $L_\sigma$ per layer can be paired
into \emph{excited states} with subsequently larger angular momenta, up to a maximum $L_\text{max}=2 L_\sigma$. For the system sizes with 
partially filled composite fermion shells in each layer, one obtains the values of $L_\sigma=\frac{3}{2}$ for $N=10$, $L_\sigma=\frac{5}{2}$ for $N=14$, 
and $L_\sigma=\frac{3}{2}+\frac{5}{2}=4$ for $N=16$ particles.
We therefore expect a low-lying mode of excitations with angular momenta up to $L_\text{max}=3$, $L_\text{max}=5$, and $L_\text{max}=8$, respectively. 
Indeed, upon inspection of the spectra, such a mode can be identified.

\begin{figure*}[tttt]
  \begin{center}
    \includegraphics[width=0.94\textwidth]{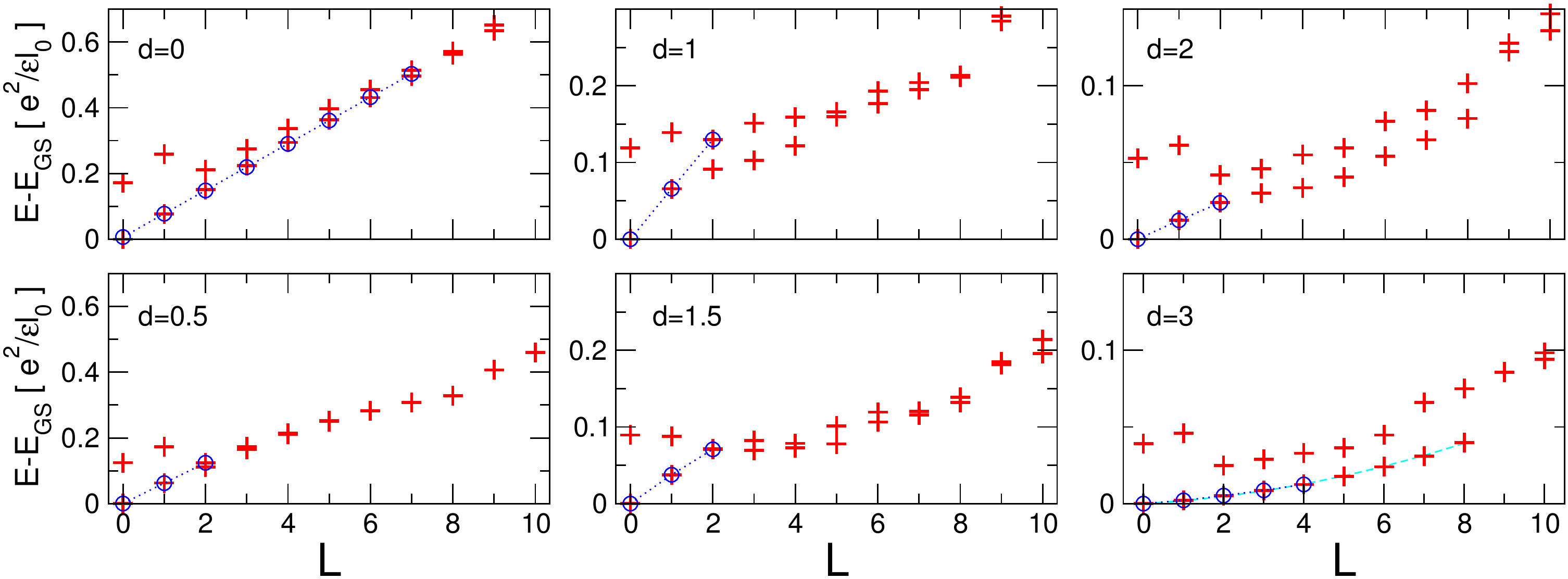}
  \end{center}
\caption{  \label{fig:spectra} Spectra of the Coulomb Hamiltonian in the bilayer system on the sphere for $N=16$ particles with $N_\phi=15$ flux, for layer
separations $d$ ranging from $d=0$ (top left) to $d=3\ell_0$ (bottom right). Energies are indicated in units of $e^2/\epsilon \ell_0$ relative to the groundstate
energy $E_\text{GS}$ at $L=0$. Red crosses mark the lowest two eigenstates in each sector of given $L$. Blue circles indicate those states which were identified
as part
of the Goldstone branch by high overlaps (see Fig.~\ref{fig:overlaps}). At $d=0$ the Goldstone mode is very well formed, while it is clearly visible at small and
intermediate $d$ how there are level crossings with additional low lying excitations. At $d=3$, a mode of low-lying excitations is once again clearly separated from
the rest of the spectrum, which terminates sharply at $L=8$. Its dispersion is linear at small $L$, but it has a quadratic component responsible of the upturn at larger $L$.
Note the change in the scale of the $y$-axis for the three columns of panels.
}
\end{figure*}

As an example, Fig.~\ref{fig:spectra} displays the spectra for the system with $N=16$ particles, including different values of the layer separation.
Indeed for $d=3\ell_0$, shown in the bottom right panel, there is clear evidence of a mode of excitations terminating at $L=8$. Its dispersion is approximately
linear at small $L$, however, it has a quadratic component as well. This should be compared to the Goldstone mode at $d=0$ (top left panel), for which linear
dispersion is clearly realized up to high values of angular momentum. 

%To distinguish the low-lying mode at large $d$, we shall name it a pseudo-Goldstone-mode.

Once the termination of the low energy branch at $L_\text{max}=8$ is identified at large $d$, it becomes apparent that this feature of a jump in the spectrum at $L_\text{max}$
exists at all values of layer separation shown, with the exception of the SU$(2)$ invariant case of vanishing $d$. Note that this termination is a feature of composite
\emph{fermion} physics, explained by successively filling the lowest shells of these composite particles, while obeying Hund's rule. 
The observation that composite fermion physics intervenes at
very small layer separation had been made previously by the current authors. While the $111$-state can be regarded as a condensate of composite \emph{bosons},
it was shown that an accurate description of the groundstate requires a mixed-fluid description of both composite bosons and composite fermions at any finite
layer separation \cite{MollerSimonRezayiPRB,PRLSimon03}. The identified jump may constitute evidence for the mixed-fluid picture 
in the excitation spectrum, but more study of these excitation will certainly be required.

A linearly dispersing mode at small $k$ exists at all values of the layer separation. The states which were shown in Fig.~\ref{fig:overlaps} to have large overlap
with the trial states (\ref{eq:TrialState}) are highlighted by blue circles in the spectra shown in Fig.~\ref{fig:spectra}. These states very accurately come to lie
on a single line, which is true especially for the $111$-state at $d=0$, but also for the intermediate layer separations such as $d=1 \ell_0$, where the first excited
state at $L=2$ lies in the continuation of the line through the points at $L=0$ and $L=1$ and is shown to be associated to the Goldstone mode by its overlap.
Judging by the spectra, it is also very suggestive that multiple level crossings occur at larger values of $L$, for example at $L=2$ for $d=1.5\ell_0$.
Finally, between $d=2\ell_0$ and $d=3\ell_0$, a change occurs in the association of the low-lying mode with the Goldstone mode trial states. For
$d=2\ell_0$ only the states up to $L=2$ have a good overlap, and the remaining states of the already well formed band of low-lying states in the exact spectrum
are of a different nature. The transition to low overlaps occurs at a point where this band has a visible kink and flattens out.
Finally at $d=3\ell_0$, this low lying band has good overlaps up to much higher momenta, which, as we have discussed above, is a reflection of two approximately uncoupled composite fermion Fermi seas which each maximizes its own angular momentum according to Hund's rule.

\begin{figure*}[ttbp]
  \begin{center}
    \includegraphics[width=0.71\textwidth]{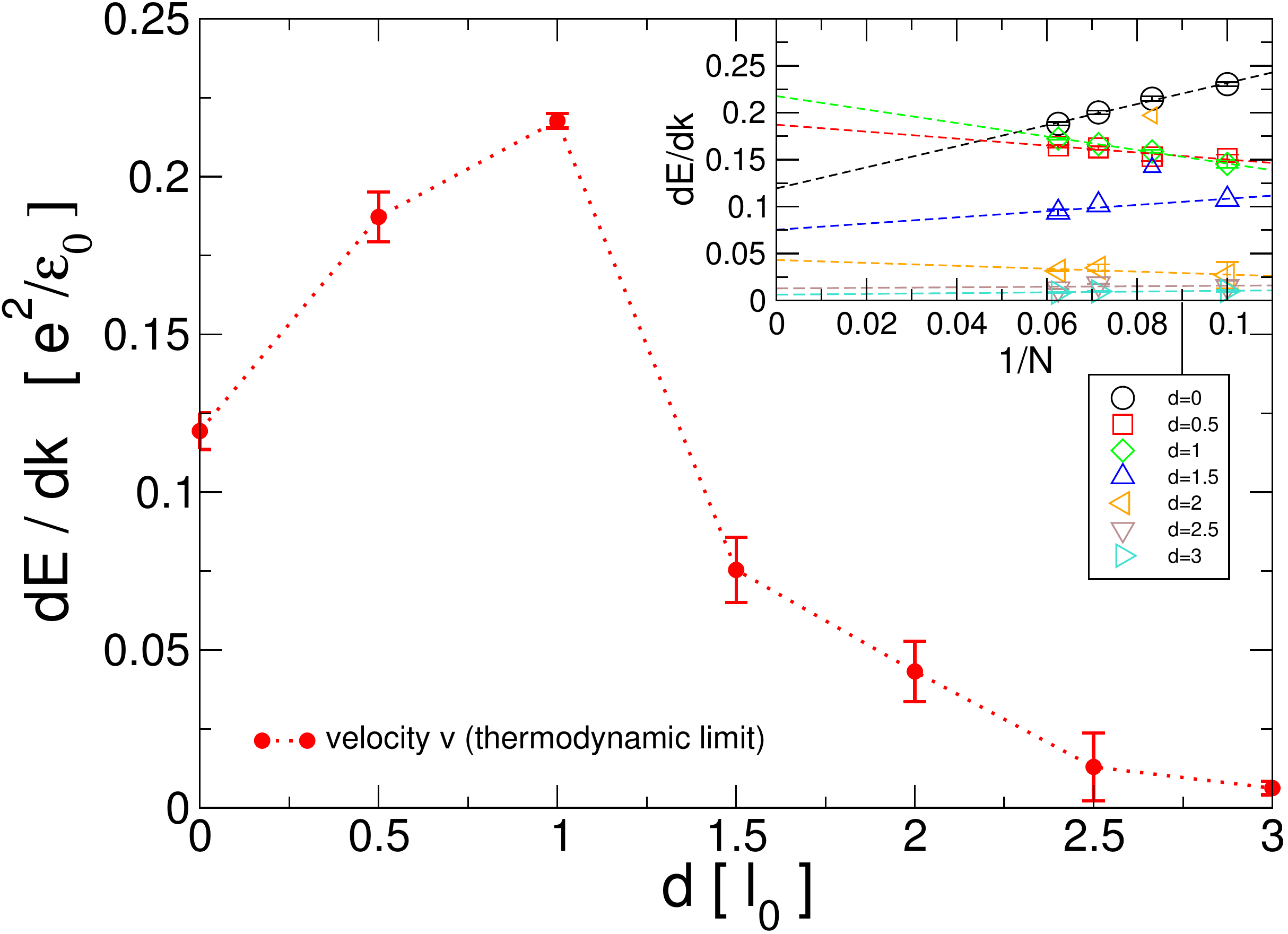}
  \end{center}
\caption{  \label{fig:velocities} The velocity of the Goldstone mode depends strongly on the layer separation, and peaks near $d\sim\ell_0$.
The velocity has significant finite-size effects on the sphere, therefore we extract an estimate of its value in the thermodynamic limit by extrapolating
the finite size values over the inverse system size (see inset).
}
\end{figure*}

Given the existence of a linearly dispersing mode over the whole range of layer separations, we now consider how its velocity changes with $d$. Comparing
the results obtained for different system sizes, a relatively strong dependence of the velocity $v=\partial E / \partial k$ is evident. Applying charging corrections
to take account of the shift in the charge-density of the system \cite{AmbrumenilMorf89} does not suffice to absorb these effects. Thus, in addition to
measuring energies in units of the rescaled magnetic length $\ell_0'=\sqrt{\nu N_\phi/N}\ell_0$, we analyze the scaling of the mode velocity as a function
of the inverse system size $N^{-1}$. These scalings are fitted well by linear extrapolation to the thermodynamic limit, as shown in the inset of Fig.~\ref{fig:velocities}.
The data for $N=12$ and large layer separation are easily identified as outliers, due to the shell filling effects discussed above. Generally, the slope is decreasing
with system size. However, at small finite layer separations the opposite scaling takes effect. The resulting estimates for the mode velocity in the thermodynamic
limit shown in the main graph of Fig.~\ref{fig:velocities} therefore show a pronounced maximum near $d=\ell_0$, which has about twice the magnitude as the value
found at $d=0$. Beyond this point, the mode velocity decreases monotonically and drops to about 1/10$^\text{th}$ the value of the $111$-state at layer separation
$d=3\ell_0$.

A previous experiment \cite{SpielmanetalLinear} has probed the velocity of the neutral mode at layer separations near the transition into the incompressible 
phase at large $d$, obtaining a combined best fit of $v=1.4 \times 10^4$ m/s for data at the three layer separations $d_1=1.61\ell_0$, 
$d_2=1.71\ell_0$ and $d_3=1.76\ell_0$. Based on linear extrapolation between our numerical data at $d=1.5\ell_0$ and $d=2\ell_0$, 
the corresponding estimates are $v(d_1)=1.14 \times 10^4$ m/s, $v(d_2)=1.21 \times 10^4$ m/s, and $v(d_3)=1.33 \times 10^4$ m/s, all slightly
smaller but within about 20\% of the proposed fit to the experimentally obtained values. Had the data in Ref.~\onlinecite{SpielmanetalLinear} been fitted 
separately at each layer separation, the velocity at $d=1.71\ell_0$ would be estimated to be about 10\% smaller than that at $d=1.61\ell_0$, roughly
reflecting the ratio of our predicted values. The data at $d=1.76\ell_0$ appears to be rather noisier, probably due to the vicinity to the phase transition, 
and would be difficult to fit on its own. We suggest that a significant enhancement of the linear mode velocity should be seen deeper inside the 
inter-layer coherent phase at smaller layer separation.

To summarize our results, we use the ansatz Eq. \ref{eq:TrialState} to construct trial wavefunctions for a low energy branch of excitations based on the 
exact ground state wavefunction.    This ansatz is accurate at all interlayer spacings $d$ when $k$ is small, and it is accurate at all $k$ when either $d$ 
is small or $d$ is large 
(so long as we do not have a filled shell configuration, whereupon only $k=0$ is in this low energy branch).   We find hints that certain aspects of the composite 
fermion physics persist to low $d$ whereas certain aspects of the Goldstone mode physics persist to high $d$.   Applying these results to the analysis of our 
numerical data we show nonmonotonic behavior of the Goldstone mode velocity as a function of the layer separation $d$. It would be interesting to look for 
this nonmonotonicity of the Goldstone mode velocity experimentally.

GM acknowledges support from Trinity Hall Cambridge, as well as from the Institute of Complex Adaptive Matter (ICAM).
SHS thanks the Aspen Center for Physics for its hospitality.

\appendix

\section{Appendix: Angular Momentum of Two Coupled Subsystems}

In this appendix, we will use the standard angular momentum notation $|l,m\rangle$ to indicate eigenstates of the $L^2$ and $L_z$ operators.

We consider two subsystems $\uparrow$ and $\downarrow$ with corresponding angular momentum operators ${\bf L}_\uparrow$ and ${\bf L}_\downarrow$.  These two subsystems combine to form the total system with angular momentum operator 
\begin{equation}
\label{eq:ang1}
\bf L = {\bf L}_\uparrow + {\bf L}_\downarrow
\end{equation}
Our objective is to show that given an eigenstate of the total system with $|l=0,m=0\rangle$ application of $(L_{+\uparrow})^J$ to this system will produce an eigenstate of the total system with $|l=J,m=J\rangle$.   To achieve this, it is sufficient to show that
$$
L_{+ \uparrow} |J, J \rangle \sim |J+1, J+1 \rangle
$$
Obviously applying $L_{+ \uparrow}$ to any state increments its overall $L_z$ eigenvalue by one (or kills the state), so all that remains is to show that applying $L_{+ \uparrow}$ to $|J, J\rangle$ results in an eigenstate $L=J+1$ of $L^2$, i.e, results in an eigenvalue of $L^2$ being given by $(J+1)(J+2)$.

Using Eq.~\ref{eq:ang1} it is just a matter of some algebra to show that
\begin{eqnarray} \nonumber
 [L_{+\uparrow}, L^2 ] &=& 2 L_{z\uparrow} L_{+\downarrow} - 2 L_{+\uparrow} L_{z\downarrow}) \\
\nonumber &=&  2  (L_+ - L_{+\uparrow})(L_z + 1)  - 2 L_{z\downarrow} L_+
\end{eqnarray}
We then apply both sides of this equation to the state $|J, J\rangle$.  Noting that $L_+$ kills $|J,J\rangle$, we obtain
\begin{eqnarray} \nonumber
L^2  [ L_{+\uparrow} |J, J\rangle ]  &=& [L_{+\uparrow} (J(J+1))  + 2 L_{+\uparrow}(J+1)] |J, J \rangle  \\
  &=& (J+1)(J+2) [ L_{+\uparrow} |J, J\rangle ]  \nonumber
\end{eqnarray}
which completes the proof.

\bibliography{Goldstone}

\end{document}